\documentclass[a4paper]{jpconf}
\usepackage{graphicx}
\begin{document}
\title{PANDA Barrel DIRC: From Design to Component Production}

\author{G~Schepers$^1$,
A~Belias$^1$,
R~Dzhygadlo$^1$,
A~Gerhardt$^1$,
D~Lehmann$^1$,
K~Peters$^{1,2}$,
C~Schwarz$^1$,
J~Schwiening$^1$,
M~Traxler$^1$,
L~Schmitt$^3$,
M~B\"{o}hm$^4$,
S~Krauss$^4$,
A~Lehmann$^4$,
D~Miehling$^4$,
M~Pfaffinger$^4$,
M~D\"{u}ren$^5$,
E~Etzelm\"{u}ller$^5$,
K~F\"{o}hl$^5$,
A~Hayrapetyan$^5$,
﻿﻿I~K\"{o}seoglu$^{5,1}$,
M~Schmidt$^5$,
T~Wasem$^5$,
C~Sfienti$^6$, 
A~Ali$^{7}$,
A~Barnyakov$^8$,
K~Beloborodov$^8$,
S~Kononov$^8$ and
I~Kuyanov$^8$}
\address{$^1$GSI Helmholtzzentrum f\"ur Schwerionenforschung GmbH, Darmstadt, Germany}
﻿﻿\address{$^2$Goethe University, Frankfurt a.M., Germany}
\address{$^3$FAIR, Facility for Antiproton and Ion Research in Europe, Darmstadt, Germany}
\address{$^4$Friedrich Alexander-University of Erlangen-Nuremberg, Erlangen, Germany}
\address{$^5$II. Physikalisches Institut, Justus Liebig-University of Giessen, Giessen, Germany}
\address{$^6$Institut f\"{u}r Kernphysik, Johannes Gutenberg-University of Mainz, Mainz, Germany}
\address{$^7$Helmholtz-Institut Mainz, Germany}
\address{﻿﻿$^8$Budker Institute of Nuclear Physics of Siberian Branch Russian Academy of Sciences, Novosibirsk, Russia}

\ead{g.schepers@gsi.de}

\begin{abstract}
Excellent particle identification (PID) will be essential for the PANDA experiment at FAIR. The Barrel DIRC  will separate kaons and pions with at least 3~s.d. for momenta up to 3.5~GeV/c and polar angles between 22 and 140~deg.
After successful validation of the final design in the CERN PS/T9 beam line, the tendering process for the two most time- and cost-intensive items, radiator bars and MCP-PMTs, started in 2018. In Sep.~2019 Nikon was selected to build the fused silica bars and successfully completed the series production of 112~bars in Feb.~2021. Measurements of the mechanical quality of the bars were performed by Nikon and the optical quality was evaluated at GSI. In Dec. 2020, the contract for the fabrication of the MCP-PMTs was awarded to PHOTONIS and the delivery of the first-of-series MCP-PMTs is expected in July 2021.
We present the design of the PANDA Barrel DIRC as well as the status of the component series production and the result of the quality assurance measurements.
\end{abstract}
\section{Introduction}
The central particle identification detector of the PANDA experiment \cite{ref:PANDA}\cite{ref:PANDAphysics} at  FAIR (Facility for Antiproton and Ion Research) has to separate pions and kaons in the challenging environment of limited space, high event rates, and a strong solenoidal magnetic field, with 3~standard deviations (s.d.) up to 3.5~GeV/c momentum. A barrel-shaped DIRC (Detection of Internally Reflected Cherenkov light) was identified as best option.

\section{PANDA Barrel DIRC Design} \label{section:DIRC}
\subsection{Motivation}
In a DIRC the produced Cherenkov photons are reflected totally internally and guided to the photon sensors. The emission angle $\theta_C$ of the photons is kept due to the squareness and parallelism of the bar sides. The polished surfaces result in a reflectivity close to~1. The optics at the end of the bar focuses the Cherenkov ring elements on an array of pixelated photon sensors. \vspace{-1pc}
\begin{center}
\begin{figure}[h] \hspace{10pc}
\includegraphics[width=15pc,angle=270]{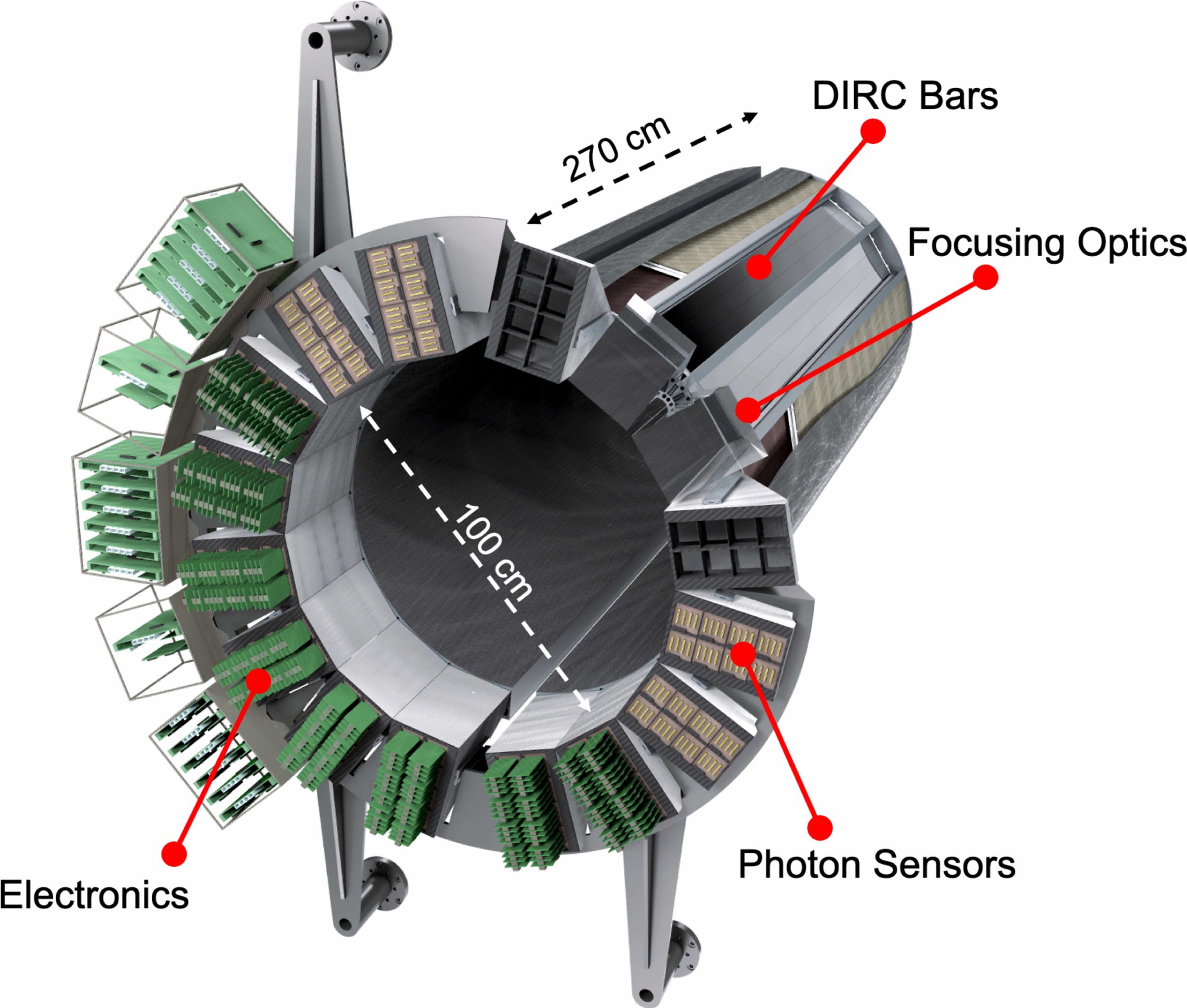}
\caption{\label{DIRC}Design of the PANDA Barrel DIRC.}
\end{figure} \vspace{-1pc}
\end{center}
Inspired by the BaBar DIRC detector \cite{ref:BaBar}, the PANDA Barrel DIRC \cite{ref:tdr} (Fig.\ref{DIRC}) was improved in several key components. The 48 bars, each 53~mm wide and 2400~mm long, are grouped in bundles of 3 in 16~CFRP bar boxes. The bar boxes are arranged at a radius of about 50~cm from the interaction point, and cover the polar angles between 22~deg and 140~deg. Each bar box is connected to a prism-shaped expansion volume made from synthetic fused silica. A three-layer spherical lens system at the up-stream end of each bar focuses the Cherenkov-photons on a flat focal plane behind the expansion volume.  Eight MCP-PMTs with 8~x~8~pixels detect the Cherenkov photons with a spatial resolution of 6~mm~x~6~mm and a timing precision of 100~ps. This allows to reconstruct the Cherenkov-pattern with a geometric method as well as with a time based imaging method. The modular design of the 16~independent sectors simplifies the construction and maintenance of the Barrel DIRC. 
\subsection{The Bars}\label{section:Bars}
The bars for the DIRC are produced from synthetic fused silica. This material is radiation hard and can be produced, formed, and polished to the required optical properties, i.e. the transparency, shape, and surface quality, respectively. The high requirements  of the PANDA experiment for a deviation from parallelism and squareness of all angles ($\textless$~0.25~mrad) of the bars and Total Thickness Variation (TTV~$\textless$~25~$\mu m$)  are provided to conserve the Cherenkov angles of the photons. The demands for the surface roughness of the faces, sides ($\textless$~5~Å~rms) and ends ($\textless$~10~Å~rms) are provided to keep the reflection losses of the Cherenkov photons small. The bars must not have bevels which would cause photon loss. 
Between 2008 and 2018 eight vendors produced bars of different material and with different production methods which have been evaluated in the optical lab and in prototype beam tests at the T9 beam line at CERN. 
\subsection{Focusing Optics, Photon Sensors and Readout} 
Between the bar and a large prism the focusing lens system is placed. It consists of a three-layer spherical lens with a layer of NLaK33 embedded between two fused silica adapter pieces \cite{ref:lens}. The prism, made from synthetic fused silica, serves as expansion volume.

The test program for the PANDA DRIC photon readout stimulated the development of lifetime enhanced MCP-PMTs. Some vendors apply the ALD (atomic layer deposition) technique and build tubes that withstand an integrated anode charge of at least 5~C/cm$^2$~\cite{ref:sensors}, the equivalent of a lifetime of 10 years in the PANDA experiment environment at design luminosity. The MCP-PMTs detect single photons with an excellent position and time resolution, even in a magnetic field of about 1~Tesla.
Highly integrated compact backplane DiRICH FPGA-TDC front-end modules \cite {ref:DiRICH} provide precise time and puls height information of the fast MCP-PMT single photon signals.
The Quality Assurance (QA) of the MCP-PMTs regarding the collection and the quantum efficiency, the rate capability, the gain, the performance in magnetic fields, and the lifetime will be provided by the Erlangen group \cite{ref:sensorsProc}. In Dec 2020 the contract for 155 MCP-PMTs was awarded to PHOTONIS, Netherlands. The first units are expected to be delivered in July~2021.
\begin{center}
\begin{figure}[t]
\includegraphics[width=11.3pc,angle=270]{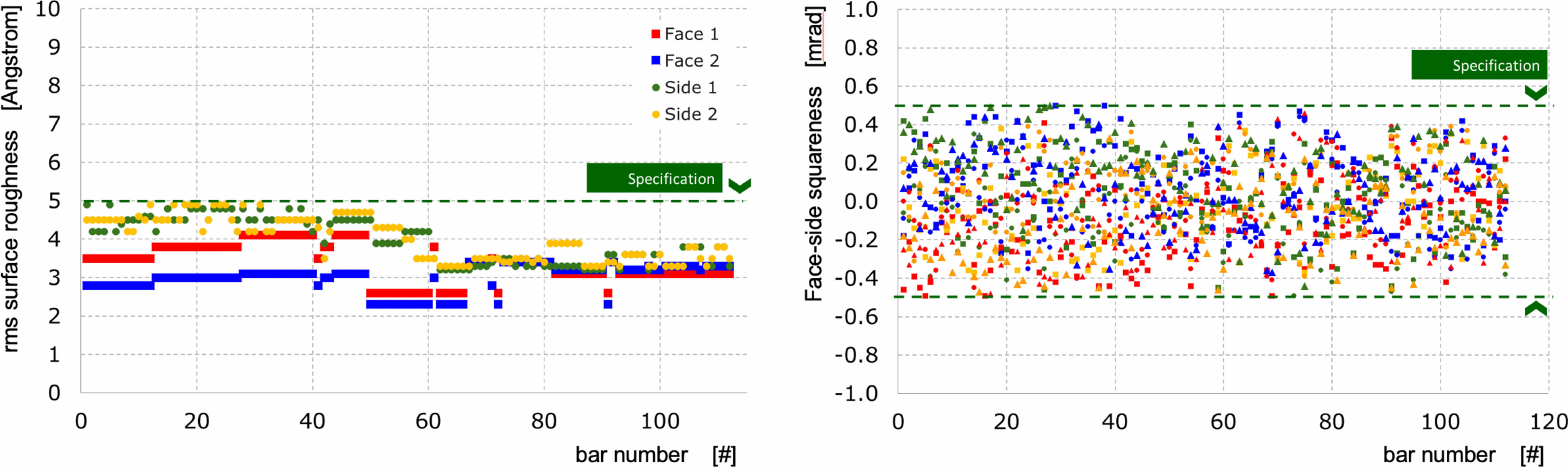}
\caption{\label{Nikon}QA-values from Nikon. Surface roughness (left) and face-side squareness (right) versus bar number.}
\end{figure}
\end{center}
\section{Series Production and QA of the DIRC Bars}\label{section:QA}
In September 2019 the contract for the delivery of the Barrel DIRC bars was awarded to Nikon Corp. Japan. The series production of DIRC bars, which has been a source of project delay for previous DIRC detectors, has proceeded smoothly and completed month ahead schedule. In February 2021 the 112 PANDA Barrel DIRC bars were completed. Nikon used their own production synthetic fused silica (NiFS-S) and an abrasive polishing method for the surface finish of the workpiece.

The in-house QA measurements of the dimensions of the bars and the material properties made by Nikon are all within the production tolerances needed for the PANDA Barrel DIRC. The distributions of the values for squareness (see Fig.\ref{Nikon} left), measured with a coordinate measuring machine (CMM), and for surface roughness (see Fig.\ref{Nikon} right), measured with a Zygo interferometer, are plotted versus the bar number. Both distributions show exemplarily the improvement of the precision with the bar production advancing in time. This process was accompanied by steady communication with the vendor. First tests at GSI revealed a deflection of a laser beam passing the bar along its long axis with and without internal reflections. Hereupon, Nikon provided additional QA values which describe the variation of the refractive index in the bar material, respectively, the shape of the bar sides. Simulations and further tests at GSI led to the conclusion that the PANDA Barrel DIRC performance is not affected by the variation of refractive index given nor the deviations from flatness of the sides of the bar. 

\begin{figure}[t]
\begin{minipage}{13pc}\vspace{5pc}
\includegraphics[width=7pc,angle=270]{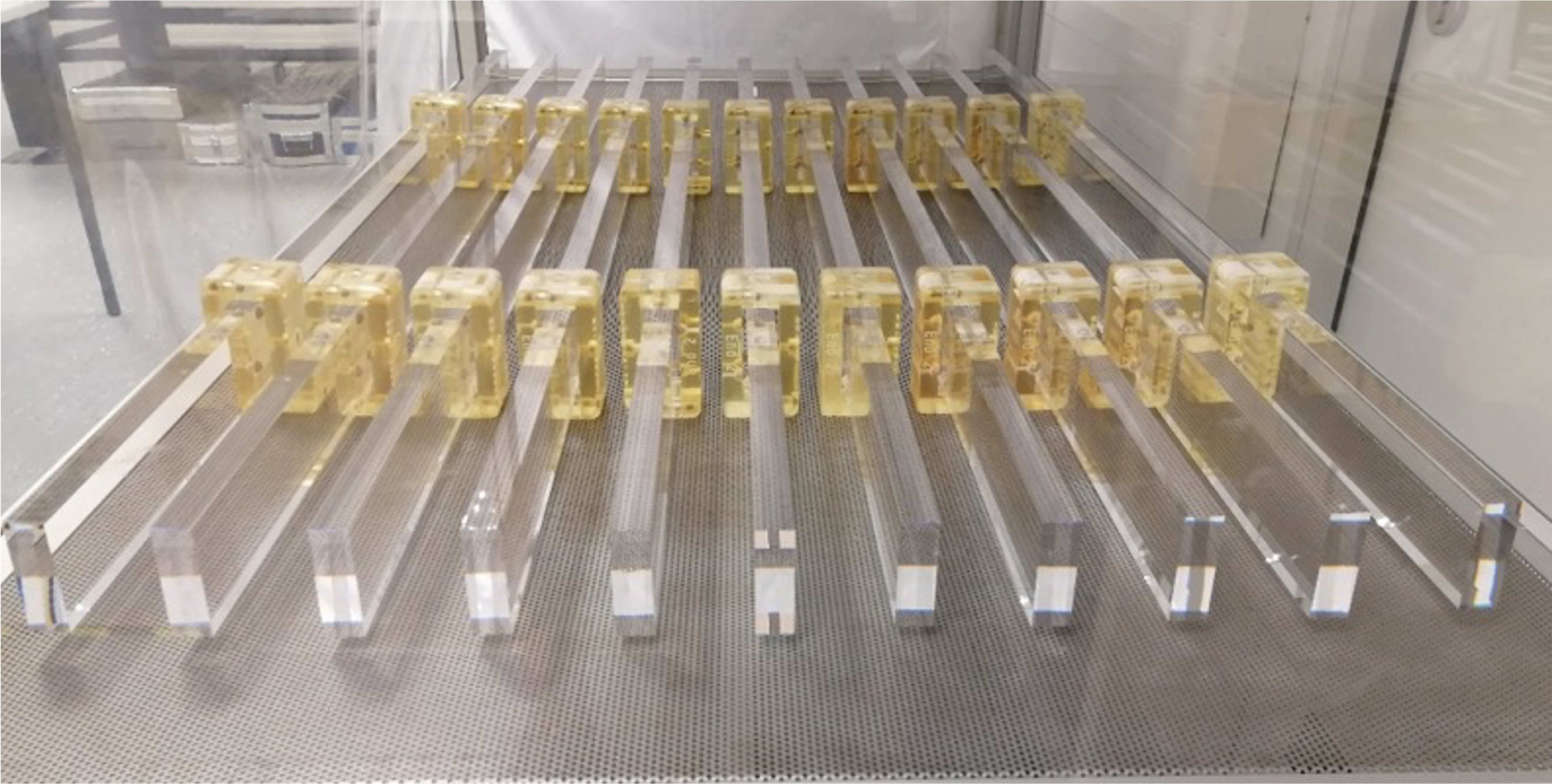}
\caption{\label{Bars}Nikon bars at GSI.}
\end{minipage}
\begin{minipage}{18pc}
\includegraphics[width=12pc,angle=270]{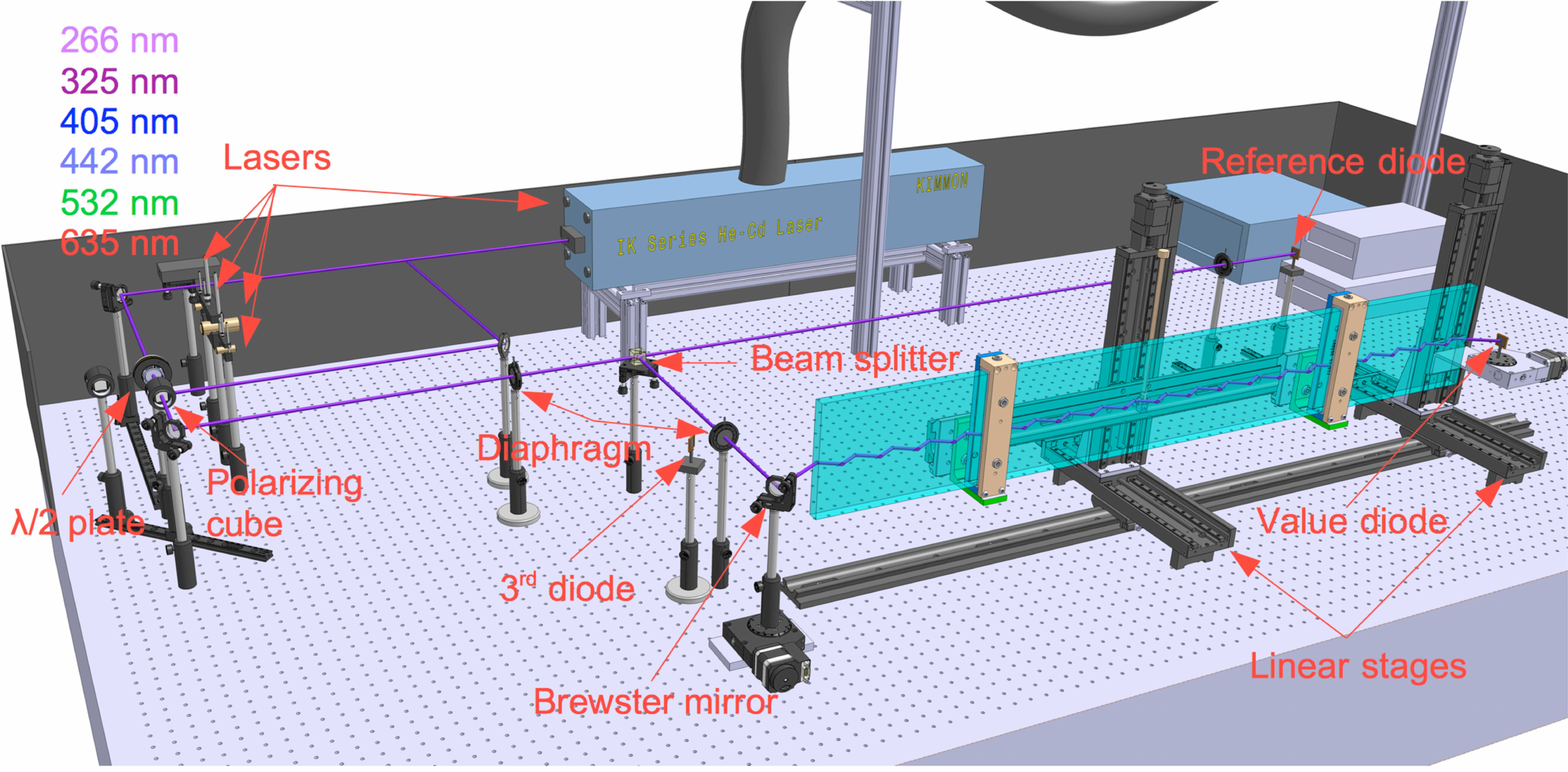}
\caption{\label{Lasersetup}Laser scanning setup at GSI.}
\end{minipage}
\end{figure}

After visible inspection the bars were installed in tagged horse shoe-like holders (see Fig.\ref{Bars}). These 3D-printed parts tag the individual bars and allow to handle and place them securely in the GSI laser scanning setup (see Fig.\ref{Lasersetup}). With the help of the setup the intensity of a polarized laser beam passing the bar under Brewster angle can be compared with the intensity of the direct beam. Thus, the reflection loss due to the about 50 internal reflections can be determined. A reference diode behind a beamsplitter serves to correct for fluctuations of the laser intensity. 
\vspace{-1pc}  
\begin{center}
\begin{figure}[h]\hspace{2.5pc}
\includegraphics[width=14.5pc,angle=270]{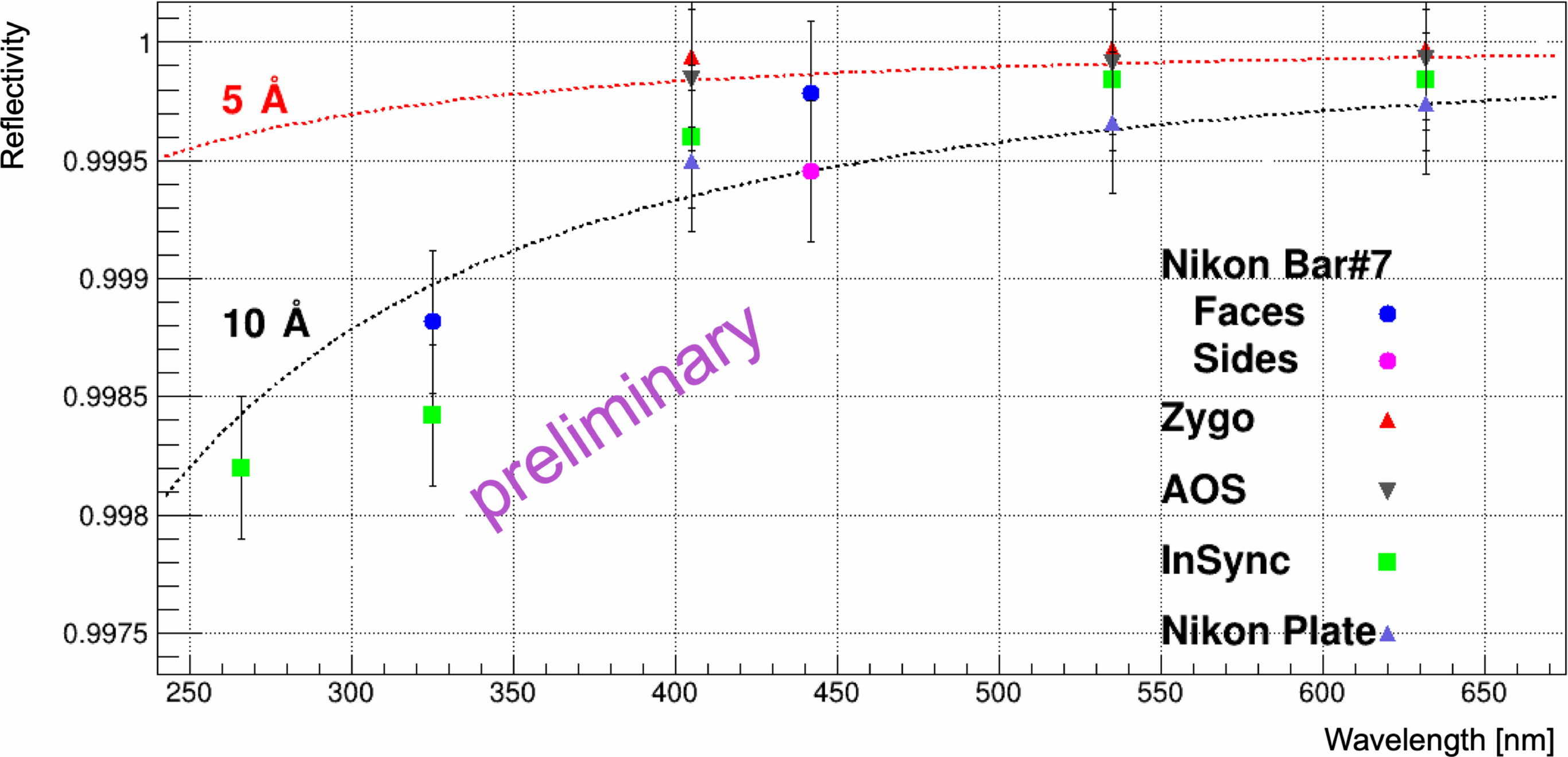}
\caption{\label{Results} Preliminary results for reflectivity compared to SST (dotted lines).}
\end{figure}\vspace{-1pc}
\end{center} 
Following the Scalar Scattering Theory (STT) \cite{ref:SST} the dependence of the reflectivity $\mathcal{R}$ from the wavelength $\lambda$ of the laser can be expressed as
\begin{math}
\mathcal{R}=1-(4\pi \cdot \mathcal{H} \cdot cos\theta \cdot n /\lambda)^2
\end{math}
with $\theta$ being the Brewster angle, $\mathcal{H}$ the internal surface roughness, and $n$ the refractive index of the material. In the laser setup the reflectivity of a bar can be determined for up to 6~wavelengths in the UV and visible light region (Fig.\ref{Results}). This allows to match the curves for SST to the results (shown for  $\mathcal{H}$~= 5~Å (red dotted line), respectively, 10~Å (black dotted line)) and to deduce the internal surface roughness with an accuracy \textless~0.5~Å. Deviations from this $1/\lambda^2$~dependence can show subsurface damage caused by the polishing method. The preliminary results for the reflectivity for one of the Nikon bars at 442~nm complies within the preliminary errors with the specified surface roughness of 5~Å. The value at 325~nm falling off the expectation gives hint to possible subsurface damage. However, Cherenkov photons with wavelengths \textless~300~nm are anyway absorbed by the epoxy, which is used to glue two bars to the total length of 2400~mm. The systematic QA of the subsurface treatment of the Nikon bars is ongoing. 
\section{Summary and Conclusion}
The PANDA Barrel DIRC project has turned from design to production. The two main components, i.e. the bars and photo sensors, have been ordered. The bars have already been produced and delivered to GSI ahead schedule. All QA values from the vendor are within specifications. The initial PANDA measurements (GSI) show no significant subsurface damage in the relevant wavelength region nor any other showstoppers. Detailed measurements of the quality of the internal surface of the Nikon bars are underway. The QA of the MCP-PMTs will start in summer 2021. Starting in 2022, the procurement process of the optics elements and bar boxes will start. The installation of the Barrel DIRC in PANDA is planned for 2025.

\end{document}